\documentclass[12pt]{article}
\title{%
The Crab Nebula: 3-dimensional Modeling}
\author{%
Shinpei SHIBATA, Haruhiko TOMATSURI, Makiko SHIMANUKI, \\
Kazuyuki SAITO, Yuji NAKAMURA, \\
{\it Department of Physics Yamagata University, Yamagata 990-8560, JAPAN, } \\
Koji MORI \\
{\it Department of Astronomy and Astrophysics, 525 Davey Laboratory, } \\
{\it The Pennsylvania State University, University Park, PA 16802, USA }}

\usepackage[dvips]{graphicx,color}
\begin{document}
\maketitle

{\small \it to appear in proc. of  "2002 International Science Symposium on
 The Universe Viewed in Gamma-Rays", 2002}

\section*{Abstract}

We construct an axisymmetric model for the Crab Nebula. The flow dyamics
is based on the model by Kennel and Coroniti (1984), but we assume
that the kinetic-energy-dominant wind is confined in the equatorial
region. We calculate the evolution of the distribution function 
of the electron-positron plasma flowing out after the shock.
Given viewing angles, we reconstruct an image of the nebula and 
also spatially resolved spectra, and we compare them with the
Chandra results.

We obtain spacially resolved spectra which agree well with the
Chandra results. However,
it is found that the assumption of toroidal field does not account
for the Chandra image. We need to assume disordered magnetic field 
with amplitude
as large as the mean toroidal field. In addition,
brightness contrast between
fore and back sides of the ring cannot be reproduced if we assumes that
the $\sigma$-parameter is as small as $\sim 10^{-3}$.
We speculate that if the magnetic energy is released by some process
making turbulent field in the nebula flow,
e.g. by magnetic reconnection, then the difficulties may be resolved:
the image can be a ring, and the
birightness contrast is higher.
The estimate of $\sigma$ can be larger than previously expected.

\section{Introduction}

A standard picture of the Crab Nebula was given by Kennel and Coroniti
(KC; 1984).
According to their picture,
a super-fast MHD wind generated by the central pulsar terminates
at a shock, and the nebula is identified as a post shock flow
shining in synchrotron radiation.
The central cavity of the nebula is occupied by the wind.
The shock position is supposed to locate at the standing inner wisp.

The KC model is very much successful because it explanes well the
synchrotron luminosity, spectrum and the size of the nebula.
An important conclusion of the KC model is that 
the energy of the wind is conveyed out not by magnetic field but by
kinetic energy in the bulk motion of the plasma.
In other words, the efficiency of
wind acceleration in the rotating magnetosphere is very high:
it is found to be 99.7\%.

The principal parameters of the pulsar wind is its luminosity
$L_w$, the Lorentz factor $\gamma_w$ of the bulk flow and
the ratio $\sigma$ of the electromagnetic energy flux to 
the kinetic energy flux,
which is referred to as the magnetization parameter.
$L_w$ is essentially the spin-down luminosity
$\approx 5 \times 10^{38}$erg/sec.
The remaining two parameters, $\gamma_w$ and $\sigma$,
together with the confining pressure $P_N$,
or equivalently the equipartition field $B_{eq} = \sqrt{4 \pi P_N}$,
determine the overall synchrotron spectrum.
Given the synchrotron luminosity
of $2 \times 10^{37}$erg/sec, the nebula size of 0.6~pc,
the peak and turn-off energy of the synchrotron spectrum of
2~eV and $10^8$~eV, respectively, it is straightfoward to obtain
the paramters:
we find
$\gamma_w =3.3 \times 10^6$,
$\sigma = 3.8 \times 10^{-3}$, and 
$B_{eq} = 0.38$mG.
One can even make an order-of-magnitude estimate to get these values
(Shibata, Kawai and Tamura 1998). 
Thus, the dominance of kinetic energy of the wind 
seems very firm.

A more rigorous fitting to the observed spectrum of the whole nebula
was made and gave similar paramters
(e.g., KC, Atoyan and Aharonian 1996).
The parameters are confirmed by observations of inverse Compton emission
in the TeV band.

The smallness of $\sigma$, in other words, dominance of the kinetic
energy is a mystery. 
Any wind theory is not able to explain how such a high
efficiency of acceleration is achieved.

Chandra observation clearly shows disc and jets and moving wisps with seed of
$0.45c$ (Mori et al. 2002), and even gives spatially resolved spectra.
Because the KC model is spherically symmetric and steady, 
the Chandra observation might seem to make the KC model useless.
However, the basic idea that the kinetic dominant wind shocks and shines
seems very firm and convincing, since 
the wind parameters explains the brightness and spectral shape of the nebula.
One may assume that the equatorial wind has different wind parameters
from the plar wind.
A latitude dependence of the wind parameters may suffice to explain
the disc-jet structure although how such a latitude dependence is made is
not known.

In this paper,
we suggest a possibility that
high spatial resolution of Chandra enables us to
examine the assumptions which were made in the KC model but have
not been checked before. In particular,
they include the ideal-MHD (frozen-in) condition
and toroidal field apprximation.
We model the nebula in 3-dimension based on the KC view and 
reproduce an image and spatially resolved spectra.
Then,
comparison with the Chandra observation (Mori 2002) is made.
In this paper, we give
preliminary results of this analysis, and 
we suggest disordered magnetic field
in the nebula. We speculate that some process which converts magnetic energy
into thermal energy, such as magnetic reconnection, 
may take place in the nebula.
If the ideal-MHD condition is broken down in the nebula, the estimated value
of $\sigma$ should be changed.

\section{A 3D Model}

Our model is based on the KC model except for that
the wind is confined within the equatorial region with width of
$\sim \pm 10^\circ$ so that the disc will be expected 
in a reproduced image (see Fig.~1).
(Althoug we also assume a polar wind by cuting the spherically symmetirc
wind out with an opeing angle of $\pm 10^\circ$ deg for the reproduction of the
image, this is just an artist's spirit,
and we do not make any analysis for the polar jets.)
\begin{figure}[t]
  \begin{center}
    \includegraphics[height=13pc]{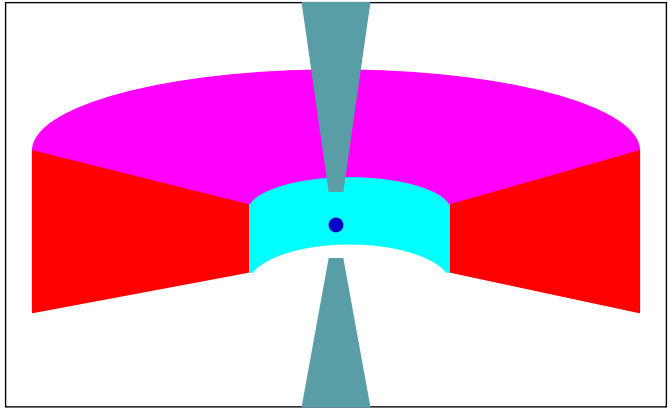}
  \end{center}
  \caption{The three-dimensional structure we assumed for the model.}
\end{figure}

In the present model,
the post shock flow (nebula flow) follows the steady solution given
by KC. To obtain this solution, 
the ideal-MHD (frozen-in) condition and
toroidal filed approximation have been used.

If $\sigma$ is much less than unity,
the speed of the flow is $\sim (1/3)c$ just after the shock
and decreases rapidly as $\propto r^{-2}$
(because the flow is subsonic, pressure is roughly uniform,
and as a result $n \approx$ const., $nr^2v \approx $const.)
Due to the deceleration, magnetic field accumulates and is amplified
according to the frozen-in condition, $B \propto r$.
Once the magnetic field increases as large as the equipartition field, 
the magnetic pressure is important in the flow dynamics.
As a result, the flow speed
satulates. This is about the region where nebula is brightest
(the typical size of the nebula is then given by $r=1/(3\sqrt{\sigma})$).
The smaller $\sigma$, the larger and brighter the nebula.
Thus, the small $\sigma$ is requried to explain the luminosity
of the synchrotron nebula as far as the ideal-MHD condition holds.
On the other hand, the flow speed is small.

The particle distribution function is assumed to be
initially a power law. Further, we assume that the distribution
function is isotropic at each point, i.e., the pitch angle distribution
is uniform.
We calculate the evolution of the distribution
function taking into account the adiabatic and synchrotron
cooling.
In order to reproduce an image, the specific synchrotron emissivity is
integrated along line-of-sight for each point on the sky.
We take into account the relativistic Doppler effect because the 
nebula flow is relativistic. 
Because the observed photons are emitted by paritcles
directed toward the observer, emissivity depends on the pitch angle of
the particles with respect to the local magnetic field.

\section{Results}

For a reproduced image
one may expect an ring since we have assume a disc wind
such as shown in the right pannel of Fig.~2.
The expected radius of the ring will be
$\sim 1/3\sqrt{\sigma}$, where the
nebula brighhtens with amplified magnetic field.
However, what we have is not a ring but is a 'lip-shaped' image
as shown in Fig.~3.
Because the pitch angles of the particles directing toward us
is small at the both corners of the ellipse, 
brightness is reduced there. 
With this effect and the central cavity, the image becomes 'lip-shaped'.
If one assumes random field with a magnitude of about the same as
the toroidal mean filed, a reproduced image becomes a ring (Fig.~2).
\begin{figure}[t]
  \begin{center}
    \includegraphics[height=15pc]{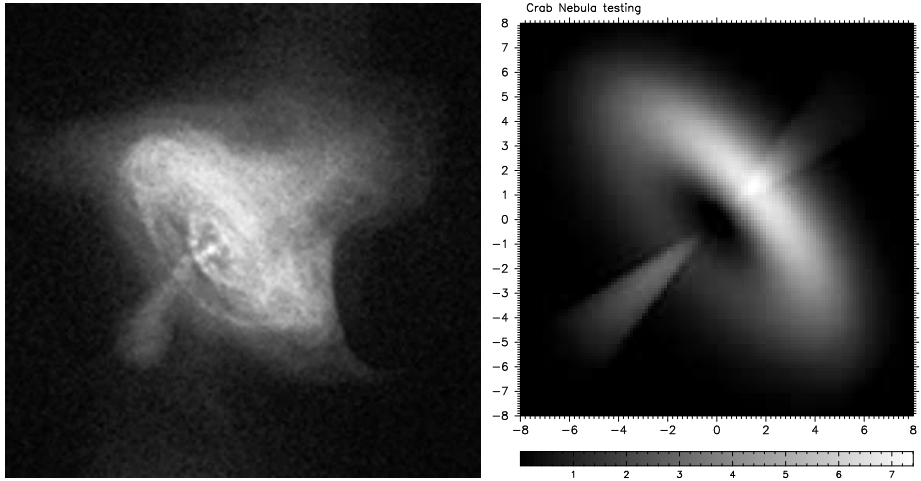}
    \includegraphics[height=15pc]{sfig3.eps}
  \end{center}
  \caption{The Chandra image (Wiesskopf 2000) and a reconstructed image
  with {\it fake parameters}, the flow speed of $\sim 0.2c$ and 
  randum field. If we follow the KC model, the flow speed is very small
  (because of the smallness of $\sigma$) and the field is domianted by
  the toroidal field. }
\end{figure}

\begin{figure}[t]
  \begin{center}
    \includegraphics[width=5.6cm]{sfig4.eps}
    \includegraphics[width=5.6cm]{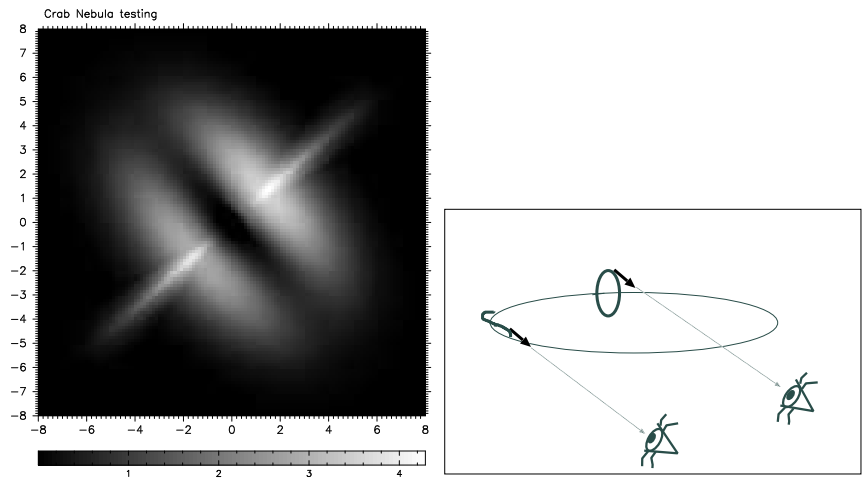}
  \end{center}
  \caption{We obtained a 'lip-shape' image (left) from our model in which
  the flow speed is given by KC, and the field is assumed to be pure
  toroidal. The dim regions in north-east and south-west (the two ends of
  the ring) is due to small pitch angels of the particles which direct toward
  us (right).}
\end{figure}

Another important thing in the reproduced image is the intensity ratio
between back and fore sides of the ring.
We obtain 1.3, but observed is $\sim 5$.
The week contrast is caused by the strong deceleration of the
nebula flow, and in turn by the smallness of $\sigma$.
As far as the intensity contrast is attributed to the Dopper boosting,
the weak construst is unavoidable in the frame work of the KC model.

We examine the 
spatially resolved spectra along a line perpendicular to the
rotation axis, which is free from the Doppler boosting effect.
According to the Chandra observation (Mori 2002),
the spectral index is about 1.7 and almost constant in the inner 
region, 
while the intensity increases with distance from the center. This continues
until synchrotron burn-off becomes evident.
Beyond a certain point, the photon index increases up to 2.5 (the spectra become
steepter), and the intensity decreases.
The result of the model calculation is 
in agreement with this Chandra observation.
However, we have to careful about the background radiation which
is caused by diffuse compontents other than the disc component.
(Detailed anaysis for this effect will be discussed in a subsequent
paper.)

\begin{figure}[tb]
  \begin{center}
    \includegraphics[height=8pc]{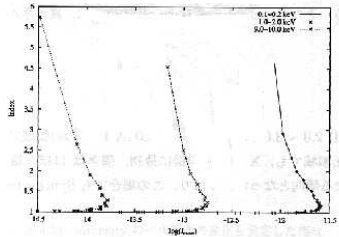}
  \end{center}
  \caption{Spectral evolution in three X-ray bands
  along a line perpendicular to the rotation axis 
  on the sky. Horizontal axis is the brightness and the vertical axis is
  the photon index. 
  As the radius increases, the nebula brightens with constant photon
  index, and at some distance, the photon index begins to increase.}
\end{figure}

\section{Conclusion}

Applying the KC model, we reconstruct an X-ray image and 
spatially resolved spectra, which are compared with the Chandra result.
The basic properties of the spectra are understood by the model.
However, the calculated image is inconsistent with the Chandra image.
Owing to the pure toroidal field,the reproduced image is
'lip-shaped'. Since $\sigma$ is asswmed to be small,
the post shock flow is decelerated,and as a result the brightress
contrast is small.
The assumptions of the toroidal field and the smallness of $\sigma$
are incompatible with the observation.
We suggest that there is significant turbulent magnetic field in
the nebula.

A model explaining the Chandra observation may be constructed 
if we assume a heating process such as magnetic reconnection
in the nebula flow. Suppose $\sigma$ is rather large and
the post shock flow is faster. The Doppler effect will cause
a higher contrast of back and fore brightness. In the nebula,
turbulent field is produced, and plasma is heated and brighten in
synchrotron radiation so that the luminosity, and the spectral shape
may be explained. 
Obliqueness of the pulsar causes a series of current sheets spaced with
the ligh cylinder radius ($\sim 10^8$cm) in the equatorial region.
This effect may account for the equatorial disc.
This is just a speculation but will be studied in detail in
a subsequent paper.

\section*{References}
\begin{verse}
Atoyan, A. M. \& Aharonian, F. A. 1996, MNRAS 278 525 

Kennel, C. F. \& Coroniti, F. V. 1984, ApJ 283 694

Kennel, C. F. \& Coroniti, F. V. 1984, ApJ 283 710

Mori, K., Hester, J.~J., 
Burrows, D.~N., Pavlov, G.~G., \& Tsunemi, H.\ 2002, ASP Conf.~Ser.~271: 
Neutron Stars in Supernova Remnants, 157 

Mori, K., 2002 PhD thesis, Ohsaka University

Wiesskopf, M. C., et al. 2000, ApJ 536 L81

\end{verse}
\end{document}